\documentclass{iau}

\usepackage{amsmath}
\usepackage{graphicx}
\usepackage{multirow}

\usepackage{natbib}
\usepackage[inline]{enumitem}

\begin{document}

\lefttitle{Johanna Hartke}
\righttitle{PNe populations in the haloes of nearby massive ETGs}

\jnlPage{1}{12}
\jnlDoiYr{2023}
\doival{10.1017/xxxxx}

\aopheadtitle{Proceedings IAU Symposium}
\volno{384}
\editors{O. De Marco, A. Zijlstra, M. Arnaboldi, R. Szczerba, eds.}

\title{Planetary nebulae populations in the haloes of nearby massive early-type galaxies}

\author{J. Hartke$^{1,2}$, M. Arnaboldi $^{3}$, O. Gerhard$^{4}$, A. I. Ennis $^{5,6}$, C. Pulsoni$^{4}$, L. Coccato$^{3}$, A. Cortesi$^{7}$, K.C. Freeman$^{8}$, K. Kuijken$^{9}$, M. Merrifield$^{10}$, N. Napolitano$^{11}$}
\affiliation{$^{1}$ Finnish Centre for Astronomy with ESO, (FINCA), University of Turku, 20014 Turku, Finland}
\affiliation{$^{2}$ Tuorla Observatory, Department of Physics and Astronomy, University of Turku, 20014 Turku, Finland}
\affiliation{$^{3}$ European Southern Observatory, Karl-Schwarzschild-Str. 2, 85748 Garching, Germany}
\affiliation{$^{4}$Max-Planck-Institut für Extraterrestrische Physik, Giessenbachstr. 1, 85748 Garching, Germany}
\affiliation{$^{5}$Perimeter Institute for Theoretical Physics, Waterloo, Ontario N2L 2Y5, Canada}
\affiliation{$^{6}$Waterloo Centre for Astrophysics, University of Waterloo, Waterloo, Ontario, N2L 3G1, Canada}
\affiliation{$^{7}$  Instituto de Fisica, Universidade Federal do Rio de Janeiro, 21941-972, Rio de Janeiro, RJ, Brazil}
\affiliation{$^{8}$ Research School of Astronomy \& Astrophysics Mount Stromlo
              Observatory, Cotter Road,
              2611 Canberra, Australia}
\affiliation{$^{9}$ Leiden Observatory, Leiden University,
              PO Box 9513, 2300~RA Leiden, The Netherlands}
\affiliation{$^{10}$ School of Physics and Astronomy,
              University of Nottingham, NG7 2RD, United Kingdom}
\affiliation{$^{11}$ School of Physics and Astronomy, Sun Yat-sen University, DaXue Road 2, 519082 Zhuhai, China}

\begin{abstract}
Planetary nebulae (PNe) are excellent tracers of the metal-poor haloes of nearby early-type galaxies. They are commonly used to trace spatial distribution and kinematics of the halo and intracluster light at distances of up to 100 Mpcs. The results on the early-type galaxy M105 in the Leo I group represent a benchmark for the quantitative analysis of halo and intragroup light. Since the Leo I group lies at just a 10 Mpc distance, it is at the ideal location to compare results from resolved stellar populations with the homogeneous constraints over a much larger field of view from the PN populations. In M105, we have -- for the first time -- established a direct link between the presence of a metal-poor halo as traced by resolved red-giant branch stars and a PN population with a high specific frequency ($\alpha$-parameter). This confirms our inferences that the high $\alpha$-parameter PN population in the outer halo of M49 in the Virgo Cluster traces the metal-poor halo and intra-group light.
\end{abstract}


\maketitle

\section{Introduction}
In the current paradigm, the evolution of early-type galaxies (ETGs) is a two-phase process \citep[e.g.][]{2009apj...699l.178n, 2010apj...725.2312o, 2016mnras.458.2371r}, where the second phase is responsible for the substantial size growth from redshift $z\approx2$ until today \citep[e.g.][]{2005apj...626..680d, 2006mnras.366..499d}. This phase is governed by numerous mergers and accretion events that are imprinted in the galaxies' haloes, where dynamical time scales are long \citep[e.g.][]{2005apj...635..931b}. Observationally, the transition from the \textit{in-situ}-dominated galaxy centres to the \textit{ex-situ}-dominated stellar haloes is signalled by radially varying light profiles \citep[morphology, colours; e.g.][]{2014MNRAS.443.1433D, 2015a&a...581a..10c, 2016apj...820...42i, 2017a&a...603a..38s} and kinematics \citep[e.g.][]{2009mnras.394.1249c, 2016mnras.457.1242f, 2016mnras.457..147f, 2018A&A...618A..94P, 2021MNRAS.504.4923D, 2022A&A...663A.135S}. However, despite the considerable efforts to observationally distinguish in-situ from ex-situ stellar populations, these observational methods do not reliably separate those populations when applied to data from cosmological simulations where the accretion histories are known \citep{2021A&A...647A..95P, 2022ApJ...935...37R}.

In dynamically complex and massive structures such as groups and clusters of galaxies, accretion processes and mergers furthermore liberate stars into the so-called intra-group (or intra-cluster) light (IGL/ICL), a diffuse stellar component that is gravitationally bound to the group (or cluster) potential \citep{1937zwicky, 1951pasp...63...61z}. As the ICL and IGL formation is intertwined with cluster and group assembly processes, characterising it can provide important constraints on the second phase of the two-phase formation scenario in dense environments. 

As \citet{2022FrASS...972283A} discuss in their recent review, there is no uniform definition facilitating IGL- and ICL-halo separation based on deep photometry, where the ICL can either be defined through multi-component surface brightness (SB) profiles \citep[e.g.][]{2005apj...618..195g, 2010apj...720..569r, 2017apj...846..139m, 2023A&A...670L..20R}, or as all the light beyond a fixed SB limit \citep[e.g.][]{2004apj...615..196f, 2005apj...631l..41m, 2018MNRAS.474..917M}. A kinematic approach to separate halo and IGL or ICL stars based on their different binding energies can be translated to observable line-of-sight velocity distributions (LOSVDs), i.e. the ensemble of stars bound to the ICL or IGL should have larger LOS velocity dispersions than the ensemble of halo stars \citep[e.g.][]{2010mnras.405.1544d, 2014mnras.437..816c}.

However, measuring kinematics from integrated-light absorption spectra in the low-SB regime is notoriously challenging. Therefore, measuring stellar kinematics in the outer haloes and surrounding IGL or ICL of ETGs relies on discrete tracers, such as planetary nebulae (PNe). Their characteristic emission-line features combined with the absence of a strong stellar continuum make PNe viable tracers of halo and ICL assembly \citep{1996apj...472..145a} up to distances of the Coma Cluster \citep[$\sim 100$\,Mpc; ][]{2005apj...621l..93g}. 

\begin{figure}
    \centering
    \includegraphics[width=0.95\textwidth]{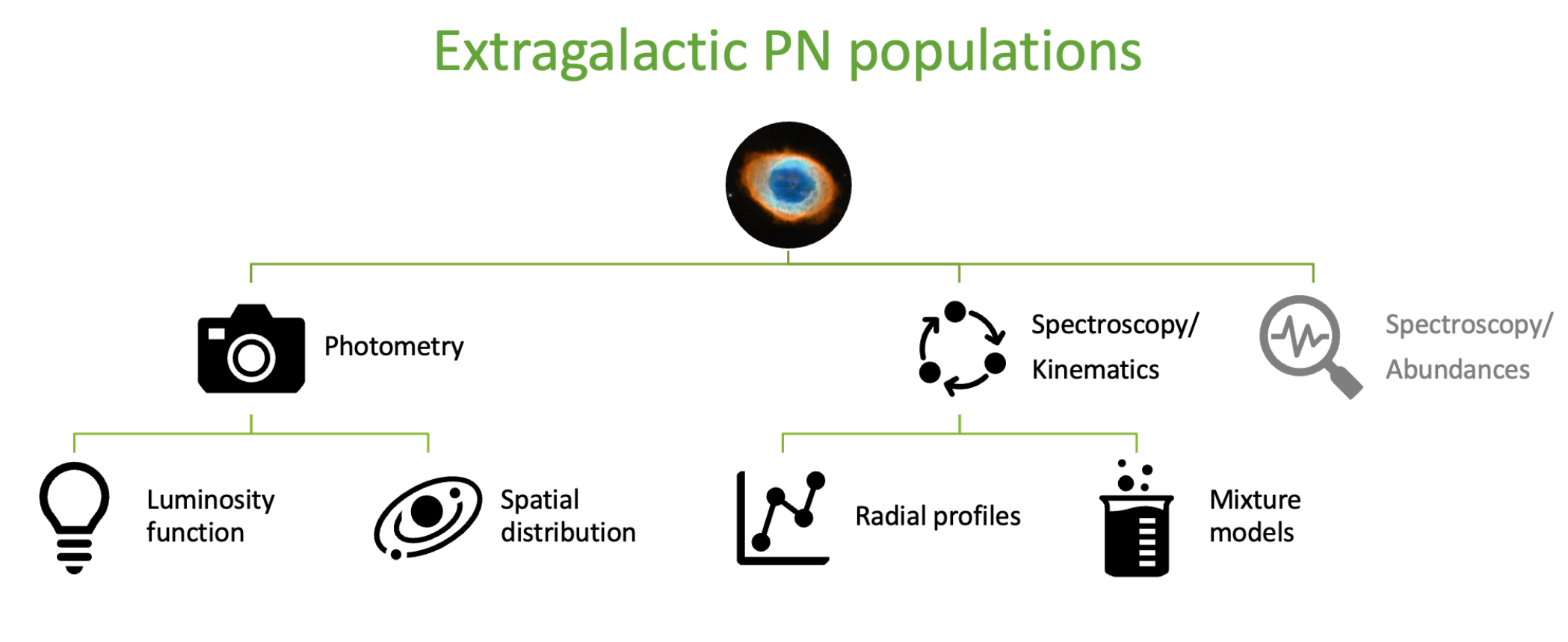}
    \caption{Overview of methods and tools used to study extragalactic PN populations. In this proceeding, we focused on photometry, i.e. deep [OIII] narrow-band surveys and spectroscopy, with the goal of obtaining the PN velocities.}
    \label{fig:Pn-are-cool}
\end{figure}

While being excellent discrete tracers of halo and ICL kinematics, PN populations may also reveal information about the underlying stellar populations. For Local Group galaxies, this can be done with direct abundance measurements (see contributions by Bhattacharya and Stanghellini in these proceedings). Still, these measurements are currently not easily accessible for galaxies at larger distances. Instead, in this proceeding, we summarise our findings on the relation of PN population properties such as the $\alpha$ parameter and PN luminosity function (PNLF) slope with the colours, ages, and metallicities of their host stellar populations. Figure~\ref{fig:Pn-are-cool} summarises the different methods and tools used to study extragalactic PN populations, especially in the halo-IGL transition context discussed here.

The $\alpha$-parameter, also known as luminosity-specific PN number, relates the number of PNe $N_\mathrm{PN}$ with the total bolometric luminosity $L_\mathrm{bol}$ of the parent stellar population \citep{1980apjs...42....1j}: 
\begin{equation}
    \alpha = \frac{N_\mathrm{PN}}{L_\mathrm{bol}}.
\end{equation}
In the context of evolving stellar populations, one can also relate the $\alpha$-parameter with the visibility lifetime $\tau_\mathrm{PN}$ of a PN population with specific evolutionary flux $\mathcal{B}$ \citep[e.g.,][]{2006mnras.368..877b}
\begin{equation}
    \alpha = \mathcal{B} \tau_\mathrm{PN}.
\end{equation}

The PNLF describes the $m_{5007}$ magnitude distribution of a PN population and can be analytically described by a combination of two exponential functions:
\begin{equation}
    N_\mathrm{PN}(m_{5007}) \propto \mathrm{e}^{c_2 (m_{5007} - \mu)}\mathrm{e}^{3(M^\star - m_{5007} + \mu)}.
    \label{eq:PNLF}
\end{equation}
The first term describes the fading brightness of a spherical gas cloud around the PN central star \citep{1963apj...137..747h}, where $c_2$ is the faint-end slope that was fixed to $c_2=0.307$ when this form of the PNLF was first introduced based on observations of M31 \citep[and other Local Group galaxies;][]{1989apj...339...53c}. This parameter was later left to vary to reconcile observations of the PNLF in the halo of the massive ETG M87 \citep{2013a&a...558a..42l}. The second term describes the bright cut-off at the absolute magnitude $M^{\star}$ for a galaxy with distance modulus $\mu$. Considering that, empirically, the absolute magnitude of the bright cut-off is near invariant with galaxy type at $M^{\star}=-4.5$ \citep{10.3389/fspas.2022.896326}, by measuring a galaxy's apparent bright cut-off $m^{\star}$ it is thus possible to infer its distance. The use of the PNLF as a secondary distance indicator is extensively discussed in the contributions by M. Roth and G. Jacoby in these proceedings. 

This proceeding is organised as follows: in Sect.~\ref{sec:surveys} we discuss the ePN.S survey of ETGs, and, building up on this, in Sect.~\ref{sec:m49} and Sect.~\ref{sec:m105}, we discuss extensions of these surveys mapping the halo-to-IGL transitions in M49 and M105 respectively. We conclude with an outlook on PN studies with current and upcoming novel instrumentation in Sect.~\ref{sec:outlook}

\section{Mapping the transition from in-situ to ex-situ stellar halo with the ePN.S survey}
\label{sec:surveys}
The extended Planetary Nebulae Spectrograph (ePN.S) ETG survey \citep{2017iaus..323..279a, 2018A&A...618A..94P} targetted 33 ETGs, of which 25 were observed with the PN.S \citep[including literature data from][]{2009mnras.394.1249c, 2013a&a...549a.115c, 2010a&a...518a..44m}, and the remaining eight ones either with other slitless spectroscopic facilities or multi-object spectrographs \citep{2005ApJ...635..290T, 2011ApJ...736...65T, 2009ApJ...691..228M, 2015A&A...574A.109W}. The survey covers a median of 5.6 effective radii ($R_\mathrm{eff}$) with a range from $3-13\;R_\mathrm{eff}$. Containing a total 8354 PNe, it is the largest kinematic survey of extragalactic PNe to date. For each of the 33 galaxies, \citet{2018A&A...618A..94P} published adaptively kernel-smoothed velocity and velocity dispersion fields,  complemented with absorption-line kinematics from the literature in the galaxies' centres. 

The ePN.S sample contains both galaxies classified as ``fast'' and as ``slow rotators'' according to the kinematics in their central regions \citep{2007MNRAS.379..401E, 2011mnras.413..813c}. As discussed in \citet{2018A&A...618A..94P}, this classification mainly pertains to processes governing the assembly of the in-situ dominated centres of ETGs. It may thus not be representative beyond the central $R_\mathrm{eff}$. In fact, \citet{2018A&A...618A..94P} found that ETGs may have very different kinematics in the halos than the slow-/fast rotator classification suggests based on the kinematics in the central regions, e.g. the onset of rotation in the halos of galaxies that were classified as slow rotators. In fact, one of the main findings of the ePN.S survey is the existence of a kinematic transition radius demarcating the inner regions and the halo. This transition is signalled by a change in the rotation velocity or the kinematic position angle. 
\citet{2018A&A...618A..94P} found that the transition radii anticorrelate with the stellar mass of the galaxies, similar to the in-situ/ex-situ transition radius found in simulations \citep{2013mnras.434.3348c, 2016mnras.458.2371r, 2022ApJ...935...37R}. However, later \citet{2021A&A...647A..95P} argue that kinematic transition radii defined by a drop in rotation or peak in the $\lambda_R$-profile\footnote{The $\lambda_R$ parameter, introduced by \citet{2007MNRAS.379..401E}, can be used to quantify whether a galaxy is rotation- or dispersion-supported. The $\lambda_R$- and specific angular momentum profiles of 32 ePN.S galaxies derived from PN.S and integral-field spectroscopic data are presented in \citet{2023A&A...674A..96P}.} 
\citep[see also][]{2020MNRAS.493.3778S} do not necessarily trace the in-situ/ex-situ transition in the IllustrisTNG ETGs. 

In the following two sections, we discuss findings from two projects that build on the ePN.S survey but have even more extended radial coverages due to additional PN.S observations and wide-field narrow-band imaging with Subaru's SuprimeCam, allowing us to uncover the kinematic transitions from the ex-situ stellar halo to the intra-group light with PNe as discrete kinematic and stellar population tracers.

\section{M49: the brightest group galaxy in the Virgo Subcluster B}
\label{sec:m49}
\begin{figure}
    \centering
    \includegraphics[width=0.5\textwidth]{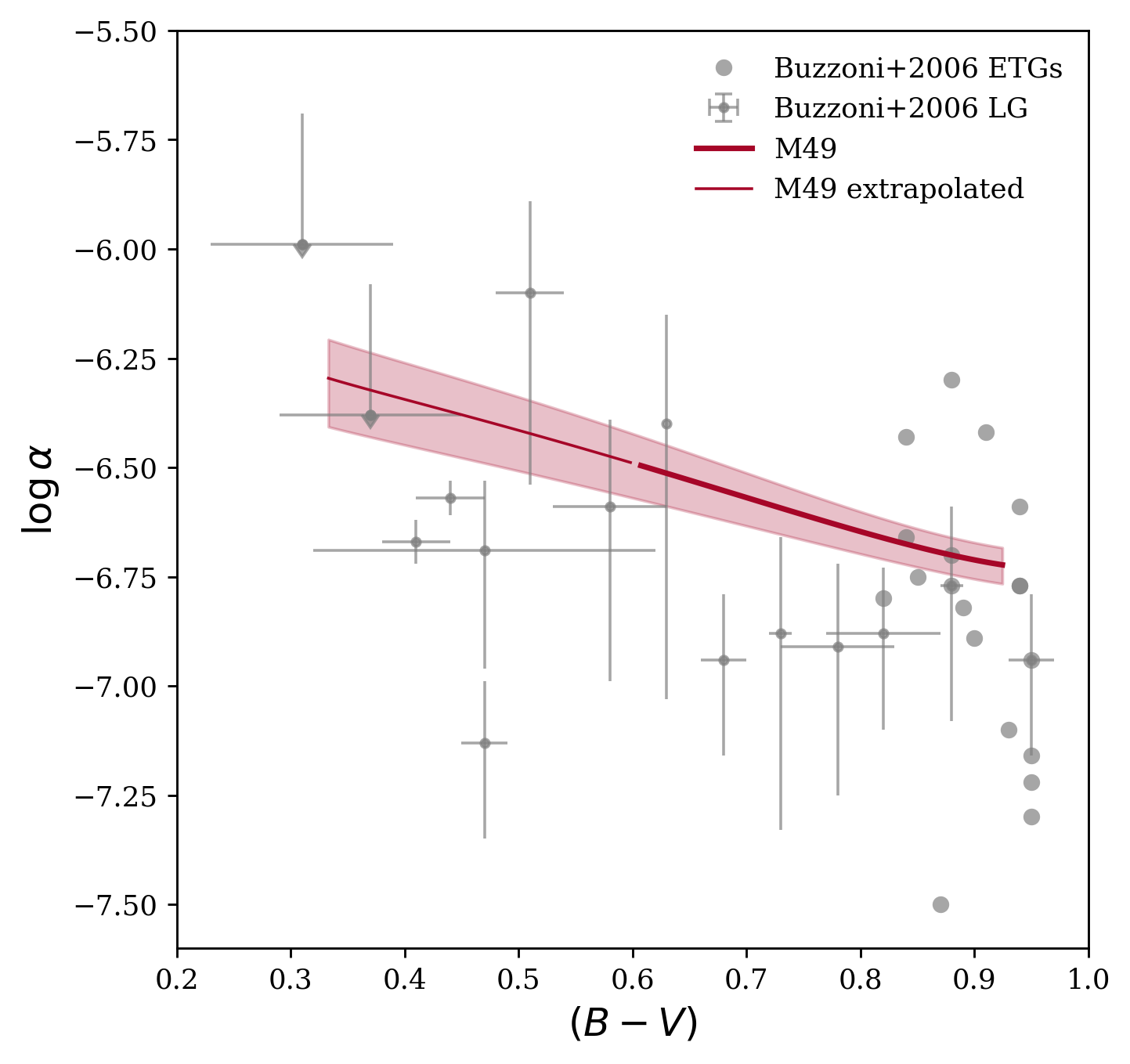}
    \caption{Variation of the M49 $\alpha$-parameter with galaxy colour \citep[solid line, colour data from][]{2013apj...764l..20m} in comparison with data from other galaxies collated in \citet[][circles and errorbars]{2006mnras.368..877b}. The colour-$\alpha$-parameter profile for M49 is based on the two-component model discussed in Sect.~\ref{sec:m49-phot} and extrapolated beyond the radial range of the colours measured by \citet{2013apj...764l..20m} assuming a linear colour gradient in the inner halo and a constant colour for the IGL. Adapted from \citet{2017a&a...603a.104h}.}
    \label{fig:M49-alpha-colour}
\end{figure}
The Virgo Cluster is the closest cluster of galaxies \citep[$D=16.5\pm0.1$,][]{2007apj...655..144m}. Its proximity has facilitated many observational campaigns that revealed a complex dynamical history. In fact, the Virgo Cluster is not virialised yet, and three subclusters/groups (A, B, and C), whose centres coincide with the massive ETGs M87, M49, and M60, were identified by \citet{1987aj.....94..251b}. First detections of intracluster PNe tracing the ICL were focused on the centre of the cluster, i.e. Virgo A, \citep{2004apj...615..196f, 2005aj....129.2585a, 2009a&a...507..621c}. Later, \citet{2013a&a...558a..42l, 2015a&a...579a.135l, 2018A&A...620A.111L} found that the Virgo A ICL has a significantly higher $\alpha$-parameter than the halo of M87 and that the change from halo to ICL is signalled not only by a tell-tale change in stellar kinematics but also by a variation in the PNLF slope (see also contribution by Arnaboldi in these proceedings).

An extended bow shock north of M49 indicates that we are currently witnessing the infall of the Virgo Subcluster B into the centre of Virgo \citep{1996apj...471..683i}. Stars that will eventually be deposited in the ICL of Virgo A are pre-processed in the Virgo Subcluster B and its IGL. Investigating the IGL in this subcluster environment may, therefore, shed light on the different processes contributing to IGL \textit{and} ICL assembly. Interestingly, the environment around M49 is characterised by a very extended halo, with a significant colour gradient, reaching blue colours of $B-V = 0.65$ at 100 kpc \citep[$\sim 6.25\;R_\mathrm{eff}$,][]{2013apj...764l..20m}.

Due to the age-metallicity degeneracy, one cannot directly infer the stellar population properties from the observed colours; therefore, the outer halo of M49 may be blue due to a relatively young (and metal-rich), or a metal-poor (and old) stellar population (or a combination of the two). A young stellar population \citep[$\sim 2$ Gyr,][]{2013apj...764l..20m} would have to be due to a recent merger with a relatively massive star-forming galaxy, but a recent merger is not visible in the deep photometry \citep{2013apj...764l..20m, 2015a&a...581a..10c, 2017apj...834...16m}. If the observed colour gradient would instead correspond to a metallicity gradient of an old stellar population (5 Gyr or older), the outermost stars would have to have metallicities below [Fe/H] $\leq -1$ \citep{2013apj...764l..20m}. If their progenitor host galaxies lie on the mass-metallicity relation \citep{2017apj...847...18z}, their masses would be of the order of $10^{-3}$ times smaller than that of M49. This is in tension with the dominant accretion channel predicted by numerical simulations, that is, mergers with mass ratios of about 1:5 \citep{2012apj...744...63o, 2016mnras.458.2371r}. To address this tension, we use PNe as discrete kinematic tracers that can help us to constrain the orbital precession time in the outer halo and IGL and thus also the dynamical ages of stars therein.

\subsection{A PN-rich IGL revealed from deep narrow-band photometry}
\label{sec:m49-phot}
Before addressing this interesting dynamical issue, by studying the distribution of PN candidates from deep Subaru SuprimeCam \citep{2002pasj...54..833m} narrow-band imaging in magnitude and position space, we could already identify indicators for a change in the PN population parameters from M49's inner to outer halo. In \citet{2017a&a...603a.104h}, we reported an excess of PNe at large radii compared to what would be expected from the $V$-band stellar light distribution, even when considering that the stellar light distribution itself deviates from a single-slope S\'{e}rsic profile at large radii \citep{2015a&a...581a..10c}. 

We could reconcile these observations with a two-component photometric model, where the inner halo is dominated by a S\'{e}rsic profile, and the light distribution at large radii is approximated by a constant surface brightness value of $\mu_{V,\mathrm{outer}} = 28.0 \,\mathrm{mag}\,\mathrm{arcsec}^{-2}$. The two model components are weighted by their respective $\alpha$-parameters. We found a 3.2 times higher $\alpha$-parameter for the constant-surface-brightness component compared to that describing the main galaxy halo. In light of the findings of \citet{2013a&a...558a..42l} for Virgo A, we interpreted this as a change of the parent stellar population from galaxy to IGL. 
In Fig.~\ref{fig:M49-alpha-colour}, we show the light-weighted $\alpha$-parameter versus colour profiles from our two-component models. As expected for massive ETGs, the $\alpha$-parameter is lowest in the inner red and metal-rich halo and then increases with decreasing colour, reaching values similar to that of star-forming local group (LG) galaxies with similarly blue colours.

While in M87 the ICL and halo PNLFs have similar faint-end slopes \citep[$c_{2,\mathrm{Halo}} = 0.72$ and $c_{2,\mathrm{ICL}} = 0.66$,][]{2015a&a...579a.135l, 2018A&A...620A.111L}, but the ICL PNLF exhibits a dip 1-1.5 magnitudes fainter than the bright cut-off, in M49 the PNLFs have significantly different faint-end slopes of $c_2 = 0.69 \pm 0.07$ in the inner and $c_2 = 1.08 \pm 0.36$ in the outer halo, also indicating a stellar population change from halo to IGL \citep{2017a&a...603a.104h}. However, these results alone are not enough to confirm the presence of IGL PNe. For that, it is necessary to measure the PN kinematics. 


\subsection{Three distinct PN populations in the halo of M49}
\begin{figure}
    \centering
    \includegraphics[width=0.7\textwidth]{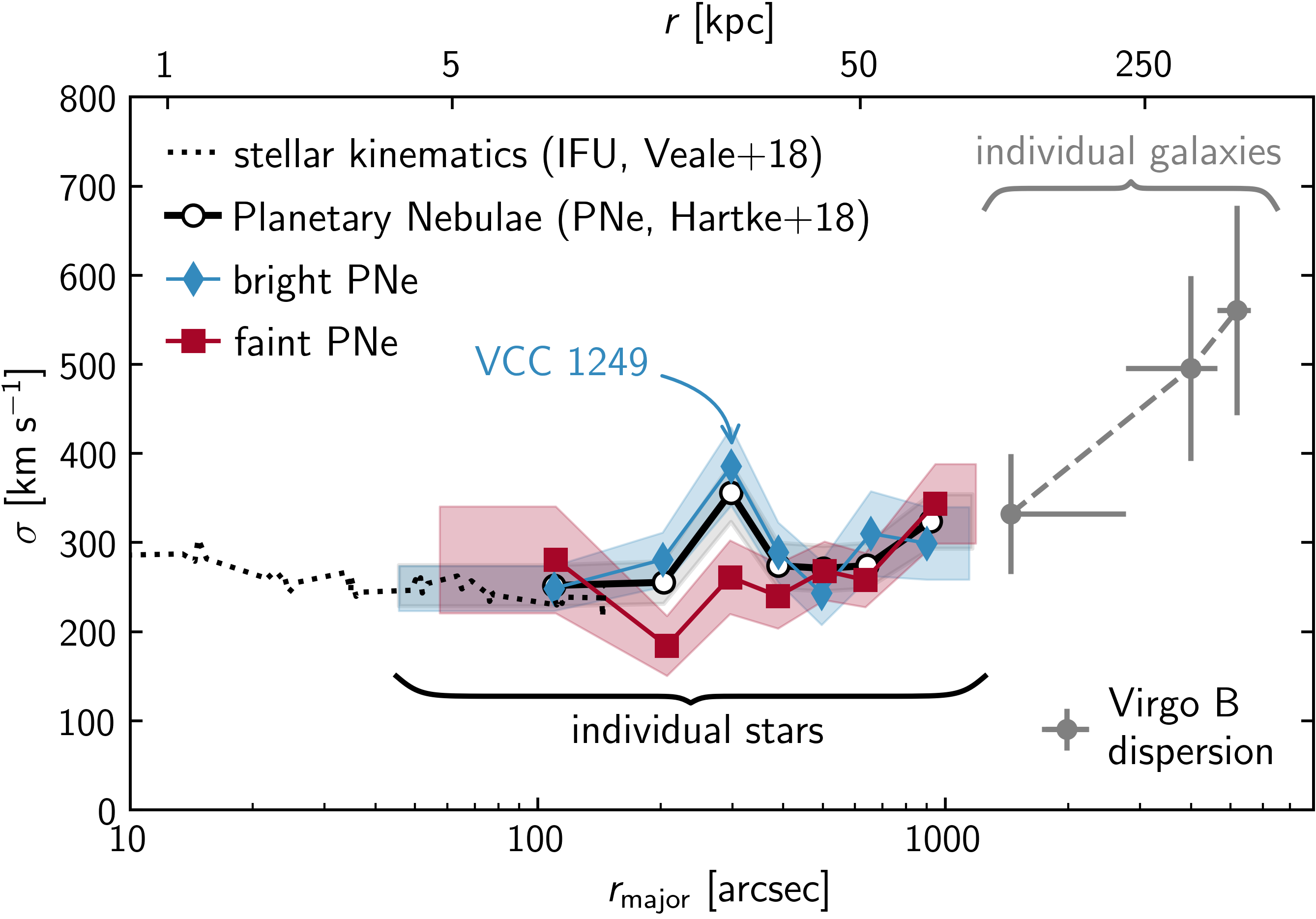}
    \caption{Line-of-sight velocity dispersion profile as a function of major-axis radius for all PNe (black), and the bright (blue) and faint (red) subsamples. Stellar kinematics derived from \citet{2018mnras.473.5446v} are denoted by the dashed line and those of dwarf galaxies in the Virgo Subcluster B \citep[from][]{2014apjs..215...22k} by grey error bars. Adapted from \citet{2018A&A...616A.123H}.}
    \label{fig:M49-sigma}
\end{figure}

Combining the data from the deep SuprimeCam narrow-band survey discussed above with that from the ePN.S survey allowed us to identify three distinct PN populations in the halo of M49 \citep{2018A&A...616A.123H}. Two of them were already discussed previously, namely the main halo of M49 and the IGL of the Virgo Subcluster B. However, the decomposition of the line-of-sight velocity distribution (LOSVD) into halo and PN components was not straightforward due to the presence of a third component, namely the dwarf irregular galaxy VCC~1249. It is interacting with M49 and dominates the blue wing of the LOSVD due to a velocity difference of $\sim 500\,\mathrm{km}\,\mathrm{s}^{-1}$ between the two galaxies. By assessing the reduced velocity as function of magnitude \citep{2006aj....131..837s} along the major and minor axes of M49, we identified a sample of bright ($m_{5007} < 27.5\,\mathrm{mag}$) PNe (258) with large reduced velocities, of which ten can be firmly associated with the interaction of M49 and VCC~1249 (see also poster by Penger in these proceedings). Their imprint on the line-of-sight velocity dispersion profile as a function of radius can be clearly seen in Fig.~\ref{fig:M49-sigma}, where the profile of the bright PNe (blue) peaks at the major-axis radius at which VCC~1249 is located. 

In the faint sample ($m_{5007} < 27.5\,\mathrm{mag}$, 178 PNe), the excess of PNe with large reduced velocities is absent and the influence of the VCC~1249 interaction negligible, facilitating the decomposition of its strongly-winged LOSVD into halo and IGL components. The transition from halo to IGL is indicated by a tell-tale rise of the line-of-sight velocity dispersion profile as a function of major-axis radius, reaching dispersion values of the Subcluster B, as illustrated in Fig.~\ref{fig:M49-sigma}. This indicates that the outermost halo of M49 is controlled by the subcluster's gravity. By using Gaussian mixture models, we find that the IGL component has a significantly larger velocity dispersion $\sigma_\mathrm{IGL} = 400\,\mathrm{km}\,\mathrm{s}^{-1}$ than the halo ($\sigma_\mathrm{halo} = 170\,\mathrm{km}\,\mathrm{s}^{-1}$). This agrees well with the expectations from simulated galaxy clusters \citep[e.g.][]{2010mnras.405.1544d, 2014mnras.437..816c}. In the faint sample, the IGL-PN fraction is about $40\%$ on average and reaches values of $57\%$ at the largest radii covered by the survey. M49 is one of the first galaxies where the transition from galaxy to cluster velocity dispersion was measured based on individual stars.

\subsection{Constraints on blue IGL formation channels in the Virgo Subcluster B from PNe}
Combining the information from deep broad- and narrow-band surveys in the Virgo Subcluster B with the dynamical information from PNe described in the previous section, it became possible to constrain the formation channels for the extended blue diffuse light first identified by \citet{2013apj...764l..20m}. In \citet{2018A&A...616A.123H}, we argued that its smooth appearance and regular, point-symmetric velocity field with large dispersion put a lower limit of $5$ Gyr on the orbital precession time, therefore ruling out a recent moderately massive merger. Instead, a viable formational channel is the early accretion of metal-poor ([Fe/H] $< -1$ and low-mass ($\sim 10^{8}\,M_\odot$) dwarf galaxies. These accretion events have much lower merger-mass ratios (about $1:10^{-4}$) than the mergers predominantly responsible for halo growth in cosmological simulations \citep[e.g.][]{2016mnras.458.2371r}. This may be due to a combination of resultion and/or feedback effects. 

Finally, having dynamically confirmed the presence of IGL PNe in the outer halo of M49, we can now also link back the high $\alpha$-parameter and steep faint-end PNLF slope, to the blue and metal-poor IGL stellar population.

\section{M105 and the IGL of the Leo~I group}
\label{sec:m105}
Compared to the Virgo Cluster, the Leo~I group in which M105 resides is a dynamically tame environment. The group's mass is two orders of magnitudes smaller \citep[e.g.][]{2017ApJ...843...16K} and at $D=10.3$ \citep{2001apj...546..681t} it is the closest galaxy group to contain both early-and late-type massive galaxies \citep{1975ApJ...202..610D}. Its proximity makes it a viable target for resolved stellar population studies of red giant branch (RGB) stars with the \textit{Hubble Space Telescope (HST)}, which revealed the emergence of a metal-poor stellar population in the outskirts of M105 based on two ACS/WFC fields at $4\,R_\mathrm{eff}$ and $12\,R_\mathrm{eff}$ respectively \citep{2007ApJ...666..903H, 2016apj...822...70l}.

\subsection{A direct link of a high $\alpha$-parameter with a metal-poor stellar population in the Leo~I Group}
\begin{figure}
    \centering
    \includegraphics[width=0.5\textwidth]{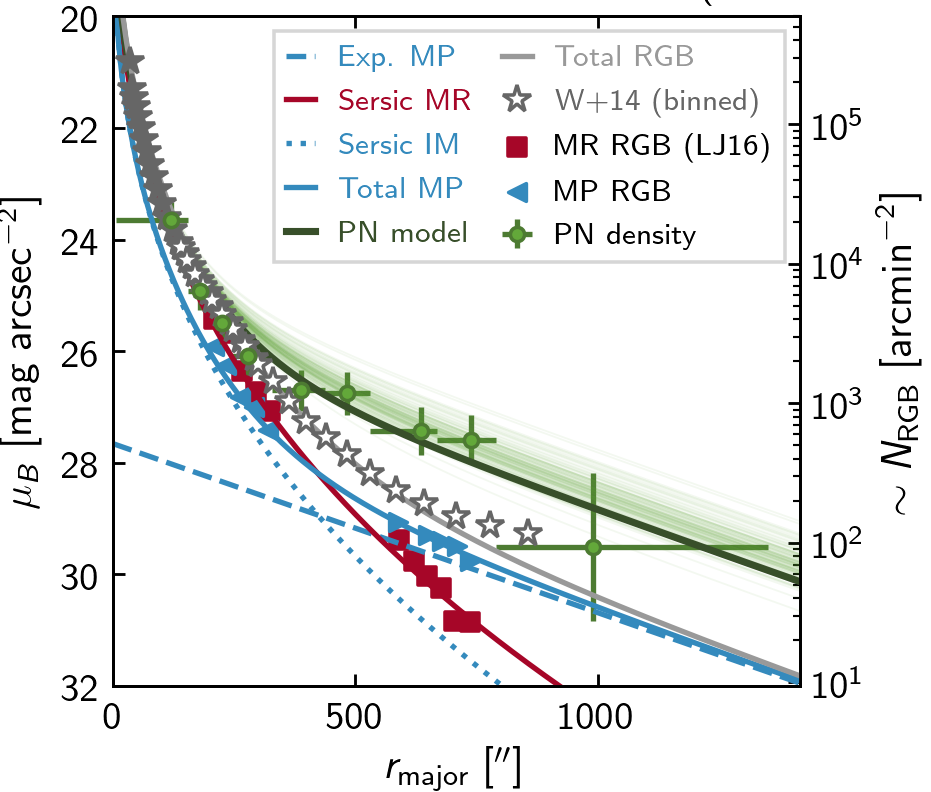}
    \caption{Observed tracer profiles and models in the halo of M105. Green dots with errorbars denote the PN number density scaled to observed $B$-band surface brightness via the $\alpha$-parameter measured in the inner halo. The best fit two-component model (including a high $\alpha$-parameter halo component, see text) and its uncertainty are indicated by the solid green line. Stars show the integrated surface brightness profile from \citet{2014apj...791...38w}, red squares [blue triangles] the number density of metal-rich [metal-poor] RGB stars from \citet{2016apj...822...70l}. The RGB number density profile (solid grey line) is decomposed into an exponential profile (dashed blue line) for the metal-poor ([M/H] $<-1$), and two S\'{e}rsic profiles for the intermediate-metallicity ($\-1\leq$[M/H]$<-0.5$, dotted blue line) and metal-rich ([M/H]$\geq-0.5$, solid red line) populations. Adapted from \citet{2020A&A...642A..46H}.}
    \label{fig:M105_density}
\end{figure}
Similar to the reasoning described in Sec.~\ref{sec:m49} for M49, we used deep SuprimeCam [OIII] narrow-band imaging covering M105 and its neighbouring galaxy NGC~3384 to study the distribution of PN candidates in magnitude and position space \citep{2020A&A...642A..46H}. As shown in Fig.~\ref{fig:M105_density}, the PN number density (green symbols and error bars) in the outer halo of the galaxy is higher than what would be expected considering a constant $\alpha$-parameter and the observed $B$-band surface brightness profile \citep[grey stars, ][]{2014apj...791...38w}. The number density distribution of metal-poor (blue triangles) and metal-rich RGB stars \citep[red squares,][]{2016apj...822...70l} can be modelled with an exponential profile for the most metal-poor stars ([M/H] $<-1$, dashed blue line) and two S\'{e}rsic profiles with the same slope for the intermediate-metallicity ($\-1\leq$[M/H]$<-0.5$, dotted blue line) and metal-rich ([M/H]$\geq-0.5$, solid red line) populations. This decomposition can be used to inform a two-component photometric model, where the inner halo is dominated by the combined S\'{e}rsic profile for the intermediate metallicity and metal-rich stellar populations and the outer halo by the exponential profile. 

The $\alpha$-parameter of the extended, metal-poor halo is more than seven times higher than that of the metal-rich inner halo. In addition to that, we also measure a steeper PNLF slope in the outer halo ($c_2 = 1.42\pm0.18$) compared to the inner halo ($c_2 = 0.51\pm0.13$). Previously, the link of an increased $\alpha$-parameter with metal-poor IGL/ICL stellar populations could only be indirectly inferred \citep{2017a&a...603a.104h, 2013a&a...558a..42l}. The origin of this stellar population, i.e. whether it was built up from pristine gas whose star-formation was truncated early as proposed by \citet{2007AJ....134...43H} or formed by accretion of metal-poor galaxies \citep{2016apj...822...70l}, is addressed in the following.

\subsection{Evidence for a kinematic transition to the metal-poor IGL}
\begin{figure}
    \centering
    \includegraphics[width=\textwidth]{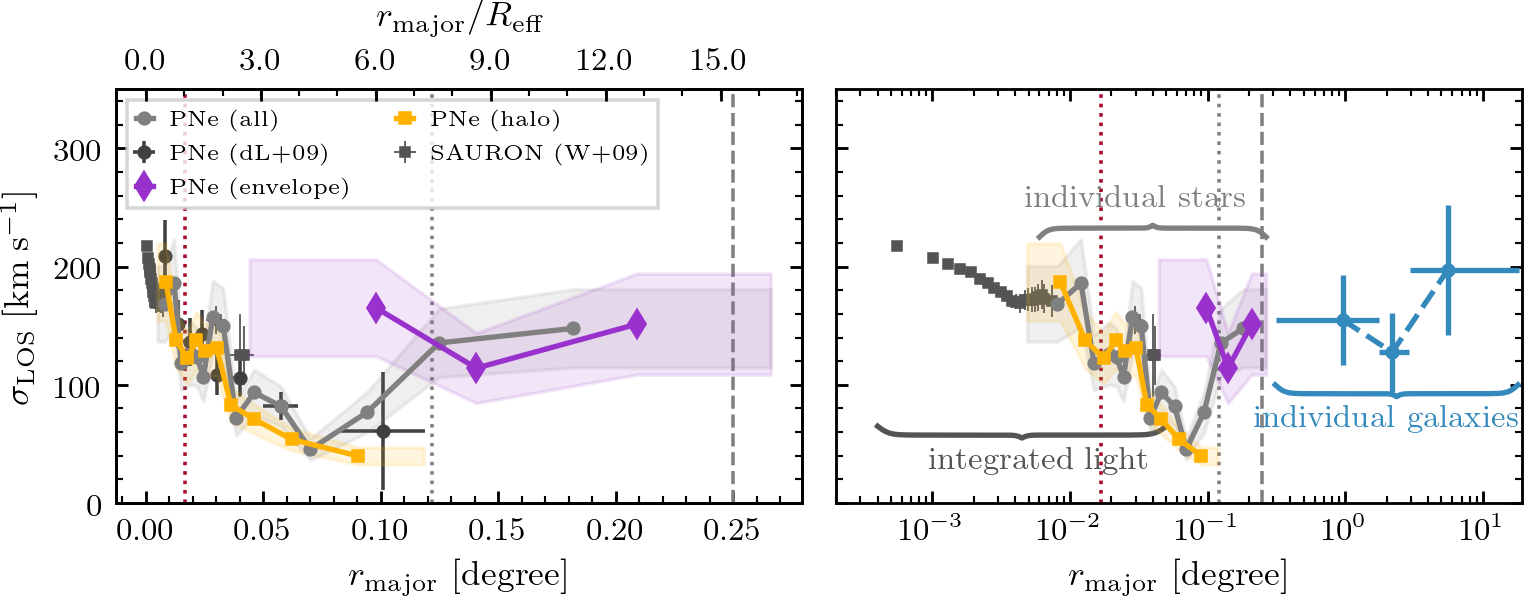}
    \caption{Line-of-sight velocity dispersion profiles in the halo of M105. Black squares denote absorption-line kinematics from \citet{2009mnras.398..561w}, black circles the PN velocity dispersion based on the central PN.S pointing \citep{2009mnras.395...76d}, grey circles the more extended data from \citet{2022A&A...663A..12H}, and purple and yellow symbols the decomposition into envelope and halo kinematics. Blue symbols on the right panel denote the dwarf galaxy kinematics derived from data compiled by \citet[][see also references therein]{2018A&A...615A.105M}. The vertical lines denote the kinematic transition from rotating core to inner halo already identified in \citet[][red dotted line]{2018A&A...618A..94P}, and from the inner halo to the exponential envelope (grey dotted). The latter transition radius agrees with the inner photometric transition radius identified independently by \citet{2022FrASS...952810R}. We also show the outer photometric transition radius that \citet{2022FrASS...952810R} interpret as the photometric transition to the IGL (dashed grey line).}
    \label{fig:M105-sigma}
\end{figure}
Early PN.S observations of the central 6 $R_\mathrm{eff}$ of M105 \citep{2003sci...301.1696r} triggered a series of papers on the dark matter (DM) content and spatial distribution of massive ETGs, as the decreasing line-of-sight velocity dispersion profile was consistent with both low-mass and low-concentration DM haloes, if any \citep{2003sci...301.1696r, 2007apj...664..257d}, as well as with DM halos in the expected mass range for massive ETGs when considering the mass-anisotropy degeneracy and viewing angle effects \citep{2005Natur.437..707D, 2007apj...664..257d, 2009mnras.395...76d, 2009mnras.398..561w}. Using new, extended kinematics of PNe allowed us to investigate whether the line-of-sight velocity dispersion profile consistent with a small DM content continued in the outer halo, as well as to constrain the formation channel of the high-$\alpha$ metal-poor stellar population discussed in the previous subsection.

The extended kinematic survey spans 15 $R_\mathrm{eff}$ and also covers the neighbouring galaxy NGC~3384, which is also a member of the Leo~I group, that was targetted as part of the PN.S survey of S0 galaxy kinematics \citep{2013a&a...549a.115c}. In \citet{2022A&A...663A..12H}, we build on the photo-kinematic decomposition techniques of \citet{2011mnras.414..642c} to determine membership probabilities for each PNe to belong to NGC~3384 or M105 and its surrounding exponential envelope. 

By studying the PN smoothed velocity fields, LOS velocity dispersion (as shown in Fig.~\ref{fig:M105-sigma}), rotation, and $\lambda_R$ profiles, we separate the PNe associated to M105 into three kinematically distinct components: 
\begin{enumerate*}
    \item the rotating core covering the central $1\,R_\mathrm{eff}$ of M105, with a metal-rich stellar population \citep[see, e.g.][]{2009mnras.398..561w} and an in-situ origin,
    \item the inner halo, whose light distribution is described by a S\'{e}rsic profile, and composed of intermediate-metallicity stars that may either have formed in-situ or were brought in through few major mergers, and
    \item the exponential envelope beyond $7.5\,R_\mathrm{eff}$, with a metal-poor and PN-rich stellar population and high velocity dispersion, reaching that of dwarf galaxies in the Leo~I group. 
\end{enumerate*}

The increase of the velocity dispersion profile at large radii, driven by PNe associated with the exponential envelope, may indicate that those stars are bound to the gravitational potential of the Leo~I group at large and thus part of its IGL. This is corroborated by the observation of the exponential envelope having similar kinematics (line-of-sight velocity dispersion, rotation amplitude, and kinematic major axis) to dwarf galaxies orbiting the Leo~I group (see right panel of Fig~\ref{fig:M105-sigma}). The transition from halo to IGL is also signalled by a break in the surface brightness profile and isophotal twists as measured by \citet{2022FrASS...952810R}. 

Our PN kinematics support the two-mode formation scenario for the metal-rich and metal-poor stellar populations in M105 and its envelope proposed by \citet{2016apj...822...70l}: the inner, metal-rich halo was likely formed in situ, while the metal-poor exponential envelope that is part of the IGL of the Leo~I group was formed via accretion. The high $\alpha$-parameter of the IGL PNe discussed in the previous subsection, similar to that of Local Group dwarf irregular galaxies \citep[see also][]{2006mnras.368..877b}, independently supports this formation channel.

\section{Outlook and perspectives for extragalactic PN studies with current and upcoming instrumentation}
\label{sec:outlook}

Our work presented above shows conclusive evidence for empirical correlations between the $\alpha$-parameter, PNLF faint-end slope $c_2$, and the underlying stellar population properties such as metallicity and colour. It builds on the seminal work of \citet{2006mnras.368..877b} on PNe of galaxy stellar populations, however, their sample of PNe and stellar population properties was derived from a heterogeneous set of observations. What is now needed is a systematic survey where PN and stellar population properties are measured simultaneously for galaxies along the entire mass-metallicity relation. However, traditional PN detection techniques, such as the on-off band technique with follow-up spectroscopic surveys or counter-dispersed imaging, are unsuited to detect PNe in the bright centres of galaxies where stellar population parameters are most easily accessible from absorption-line measurements. Integral-field spectrographs (IFSs) allow for a pixel-by-pixel decomposition of the spectra into their respective nebular and stellar (i.e. galaxy light) contributions. The SAURON IFS provided a first view of the PN population in the central regions of M31 \citep{2011mnras.415.2832s}. 

With its $1\times1$ arcmin$^{2}$ field-of-view (FoV) and large wavelength range, the Multi-Unit Spectroscopic Explorer \citep[MUSE,][]{2010SPIE.7735E..08B} at ESO's \textit{Very Large Telescope (VLT)} has opened a new window for PN discoveries in the very centres of galaxies and allows for a simultaneous measurement of stellar kinematics, ages, and metallicities. The combination of MUSE with the adaptive optics system GALACSI \citep{2008SPIE.7015E..24A, 2012SPIE.8447E..37S} can significantly boost the detection of PNe due to their point-like nature. Different strategies have been established for optimising the detection of PNe in MUSE cubes, such as 
\begin{enumerate*}
    \item the Diffuse Emission Line Filter technique \citep[DELF,][see also contribution by Soemitro in these proceedings]{2021ApJ...916...21R},
    \item using ``pure'' emission-line datacubes after subtracting the stellar continuum \citep[][see also contribution by Schlagenhauf in these proceedings]{2020A&A...637A..62S}, or 
    \item the detection based on line maps \citep{2022MNRAS.511.6087S}.
\end{enumerate*}
The methods can also be combined, as demonstrated by \citet[][these proceedings]{2022BAAA...63..175E, 2023arXiv231109176E}, who used the DELF technique on continuum-subtracted MUSE cubes. 

The imaging Fourier transform spectrometer Spectrograph SITELLE \citep[Spectromètre Imageur à Transformée de Fourier pour l'Etude en Long et en Large de raies d'Emission,][]{2019MNRAS.485.3930D} at the \textit{Canada-France-Hawaii Telescope} with its large FoV ($11\times11$ arcmin$^{2}$) and IFU capabilities in spectral filters covering some of the most important emission lines of PNe ([OIII], H$\alpha$, H$\beta$, [SII]) has the capabilities of mapping PN populations in nearby galaxies in a single pointing. First results in the scope of the SIGNALS survey \citep[][]{2019MNRAS.489.5530R} were recently published by \citet{2023MNRAS.524.3623V}.
In light of current and upcoming large surveys with IFSs, machine learning has become increasingly important to classify a large number of ionised nebulae, including, but not limited to PNe \citep{2023A&A...672A.148C, 2020ApJ...901..152R}. 

Combining IFS data with that obtained with ``traditional'' techniques in the halos of nearby galaxies may provide constraints on the variation of the $\alpha$-parameter and PNLF within a galaxy, representing the different stellar populations of in-situ and ex-situ origins, as we illustrated with the example of M105, where a combination of MUSE, \textit{HST}, and PN.S data showed a trend of a decreasing $\alpha$-parameter with metallicity.
To interpret these observations and advance our understanding of PNe as stellar population tracers, it is important that extragalactic PN populations can also be adequately modelled and compared with other stellar population properties in numerical simulations of galaxy evolution. The extension of the formalism described in \citep{2019ApJ...887...65V} to cosmological simulations, as presented at this meeting, is an important milestone.

In the coming decade(s), several upcoming instruments and facilities may open new and exciting discovery spaces for extragalactic PN science. One of them is BlueMUSE \citep{2019arXiv190601657R} at the \textit{VLT}, with an instrumental design similar to MUSE, but with a larger FoV and providing access to bluer emission lines such as [OII] at 3727 \AA, H$\gamma$ at 4340 \AA, and [OIII] at 4346 \AA, with the latter being an important ingredient in direct abundance determinations. Spectroscopic facilities with high multiplexing and IFS capabilities and unprecedentedly large FoVs, such as the Maunakea Spectroscopic Explorer \citep{2019arXiv190404907T} in the northern and the Wide-field Spectroscopic Telescope \citep{2018SPIE10700E..4EP} in the southern hemisphere at 10-m class telescopes have the potential to be transformational for extragalactic studies of PNe.

\section*{Acknowledgements}
J.H. would like to thank the organisers of IAU Symposium 384 for the invitation to present this work and for the stimulating conference. J.H. and A.I.E. acknowledge the financial support from the visitor and mobility program of the Finnish Centre for Astronomy with ESO (FINCA), funded by the Academy of Finland grant nr 306531. The authors would like to acknowledge the contribution from Dr Nigel Douglas, first P.I. of the PN.S. We also would like to thank the staff at the WHT for their support during the many PN.S observing runs.

\bibliography{literature}

\end{document}